\documentclass[12pt]{article}
\usepackage{graphicx}
\usepackage{hyperref}

\def \bea{\begin{eqnarray}}
\def \beq{\begin{equation}}
\def \ca{{\cal A}}
\def \eea{\end{eqnarray}}
\def \eeq{\end{equation}}

\def \oks{\overline{K}^{*0}}

\def \s{\sqrt{2}}
\def \st{\sqrt{3}}

\def \thet{\theta_\eta}

\def \c{\circ}

\begin{document}

\rightline{EFI 10-26}
\rightline{arXiv:1010.1770}
\rightline{October 2010}

\bigskip
\centerline{\bf CROSS RATIOS BETWEEN DALITZ PLOT AMPLITUDES}
\centerline{\bf IN THREE-BODY $D^0$ DECAYS}
\bigskip
\centerline{Bhubanjyoti Bhattacharya and Jonathan L. Rosner}
\centerline{\it Enrico Fermi Institute and Department of Physics}
\centerline{\it University of Chicago, 5640 S. Ellis Avenue, Chicago, IL 60637}

\begin{quote}
A recent study of $D^0 \to \pi^0 K^+ K^-$ and $D^0 \to K_S \pi^+\pi^-$ describes
a flavor-symmetric approach to calculate relative amplitudes and phases, for
characteristic interferences between $D$ decays to a light pseudoscalar $P$ and
a light vector $V$, on Dalitz plots for $D \to PPP$ decays.  The
flavor-symmetric approach used an earlier fit to $D \to P V$ decay rates and
was found to agree fairly well with experiments for $D^0 \to \pi^0 \pi^+ \pi^-$
but not as well for $D^0 \to \pi^0 K^+ K^-$ and $D^0 \to K_S \pi^+\pi^-$.  The
present work extends this investigation to include $D^0 \to K^- \pi^+ \pi^0$.
We use an SU(3) flavor symmetry relationship between ratios of Cabibbo-favored
(CF) $D \to P V$ amplitudes in $D^0 \to K^- \pi^+ \pi^0$ and ratios of singly-
Cabibbo-suppressed (SCS) $D \to P V$ amplitudes in $D^0 \to \pi^0 K^+ K^-$ and
$D^0 \to \pi^0 \pi^+ \pi^-$. We observe that experimental values for Dalitz
plot cross ratios obey this relationship up to discrepancies noted previously.
The need for an updated  Dalitz plot analysis of $D^0 \to K^- \pi^+ \pi^0$ is
emphasized.
\end{quote}

\leftline{PACS numbers: 13.25.Ft, 11.30.Hv, 14.40.Lb}
\bigskip

\section{Introduction}

Decays of $D$ mesons to a light pseudoscalar meson $P$ and a light vector meson
$V$ were studied earlier in Refs.\ \cite{Bhattacharya:2008ke, Cheng:2010ry}
using SU(3) flavor-symmetry. The results of these analyses were applied to
extract relative phases and amplitudes for quasi-two-body ($P V$) final states
in three-body Dalitz plots.  Agreement with experiment \cite{Aubert:2007ii,
Gaspero:2008rs, Gaspero:2010pz} was found to be good for the $D^0 \to
\pi^0 \pi^+ \pi^-$ Dalitz plot \cite{Bhattacharya:2010id}, but poorer
\cite{Bhattacharya:2010ji} for $D^0 \to \pi^0 K^+ K^-$ and $D^0 \to K_S \pi^+
\pi^-$ \cite{Aubert:2005iz, Aubert:2008bd, Poluektov:2010wz,
delAmoSanchez:2010xz, Cawlfield:2006hm, Aubert:2007dc}.

In this paper we revisit the relative phases and amplitudes of characteristic
interferences of $D \to P V$ decays on $D^0 \to P P P$ Dalitz plots. We compare
the Dalitz plot analyses of $D^0 \to \pi^0 K^+ K^-$ and $D^0 \to \pi^0 \pi^+
\pi^-$ which were previously studied in two different contexts. We also include
a comparison of these Dalitz plots with the Dalitz plot for $D^0 \to K^- \pi^+
\pi^0$.  We compare ratios of $D \to P V$ amplitudes obtained from each Dalitz
plot analysis with predictions from the flavor-symmetric technique, and find a
fair match with experimental data.  This agreement is useful in validating both
the flavor-symmetric technique and the sign conventions used in each Dalitz
plot analysis.

In Sec.\ II we recall our notation for the SU(3) flavor-symmetric analysis and
quote the values of the relevant parameters obtained in earlier fits
\cite{Bhattacharya:2008ke, Cheng:2010ry}. In Sec.\ III we construct the $D \to
P V$ amplitudes that are relevant for our present study. Sec.\ IV compares
ratios of $D \to P V$ amplitudes obtained using Dalitz plot fit fractions with
the predictions of the flavor-symmetric analysis.  We compare our results with
those of previous analyses in Sec.\ V and conclude in Sec.\ VI.

\section{Amplitudes from previous fits}

The notation for the SU(3) flavor-symmetric analysis of $D \to PV$ decays is
discussed in Ref.\ \cite{Bhattacharya:2008ke}. Here we briefly recall some of
the salient features. We denote Cabibbo-favored (CF) amplitudes, proportional
to the product $V_{ud} V^*_{cs}$ of Cabibbo-Kobayashi-Maskawa (CKM) factors, by
unprimed amplitudes. The singly-Cabibbo-suppressed (SCS) amplitudes,
proportional to the product $V_{us}V^*_{cs}$ or $V_{ud}V^* _{cd}$, are then
obtained by using the ratio SCS$/$CF$=\tan\theta_C\equiv\lambda=0.2305$
\cite{BM}, with $\theta_C$ the Cabibbo angle and signs governed by the relevant
CKM factors.

\begin{figure}
\mbox{\includegraphics[width=0.46\textwidth]{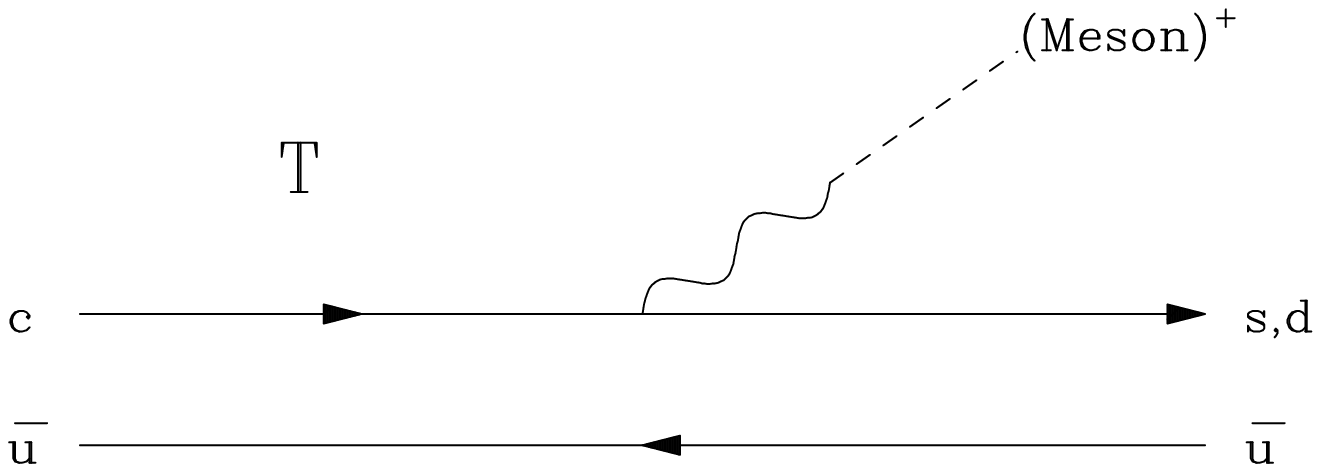} \hskip 0.3in
      \includegraphics[width=0.46\textwidth]{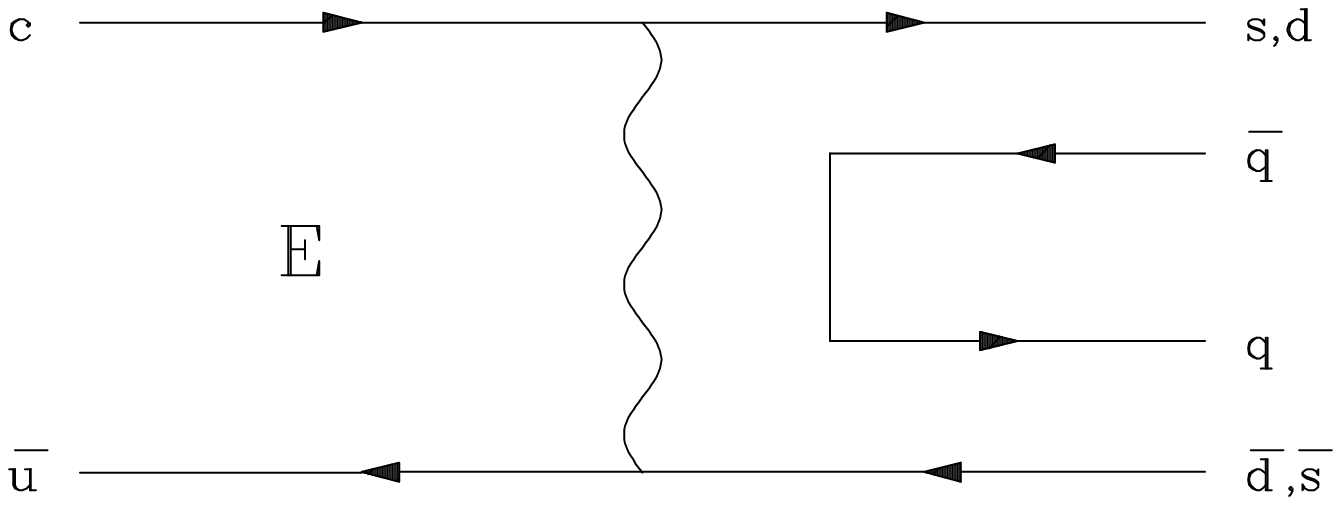}}
\caption{Flavor topologies for describing charm decays.  $T$: color-favored
tree; $E$ exchange.  Not shown:  $C$ (color-suppressed tree); $A$
(annihilation).
\label{fig:TE}}
\end{figure}

The present scenario involves the amplitudes labeled as $T$ (``tree'') and $E$
(``exchange''), illustrated in Fig.\ \ref{fig:TE}. The subscript $P$ or
$V$ on an amplitude denotes the meson ($P$ or $V$) containing the spectator
quark in the $P V$ final state. The partial width $\Gamma(H \to P V)$ for the
decay of a heavy meson $H$ is given in terms of an invariant amplitude $\ca$ as
\beq
\Gamma(H \to P V) = \frac{p^{*3}}{8\pi M^2_H}|\ca|^2,
\eeq
where $p^*$ is the center-of-mass (c.m.) 3-momentum of each final particle, and
$M_H$ is the mass of the decaying heavy meson. With this definition the
amplitudes $\ca$ are dimensionless.

The amplitudes $T_V$ and $E_P$ were obtained from fits to rates of CF $D \to
P V$ decays not involving $\eta$ or $\eta'$ \cite{Bhattacharya:2008ke}.  To
specify the amplitudes $T_P$ and $E_V$, however, one needs information on the
$\eta$--$\eta'$ mixing angle ($\thet$). Table \ref{tab:tveptpev} summarizes
these results for two values $\thet = 19.5^\circ$ and $11.7^\circ$.  In order
to make the discussion complete we also quote the results for $C_P$ and $C_V$
in Table \ref{tab:tveptpev}. As described in \cite{Bhattacharya:2008ke}
the amplitudes and phases for these parameters were also obtained from fits to
rates of CF $D \to P V$ decays.

\begin{table}[h]
\caption{Solutions for $T_V$, $E_P$, $C_P$, $T_P$, $E_V$ and $C_P$ amplitudes in
Cabibbo-favored charmed meson decays to $PV$ final states, for $\eta$--$\eta'$
mixing angles of $\thet = 19.5^\c$ and $11.7^\circ$.
\label{tab:tveptpev}}
\begin{center}
\begin{tabular}{c c c c c} \hline \hline
 & \multicolumn{2}{c}{$\thet=19.5^\c$} & \multicolumn{2}{c}{$\thet=11.7^\c$} \\
$PV$ &  Magnitude  &  Relative  &  Magnitude  &  Relative \\
ampl.& ($10^{-6}$) & strong phase & ($10^{-6}$) & strong phase \\ \hline
$T_V$& 3.95$\pm$0.07 & -- & \multicolumn{2}{c}{These results are}\\
$E_P$& 2.94$\pm$0.09 & $\delta_{E_PT_V} = (-93\pm3)^\circ$ & \multicolumn{2}{c}
{independent of $\thet$}\\
$C_P$& 4.88$\pm$0.15 & $\delta_{C_PT_V} = (-162\pm1)^\circ$ \\ \hline
$T_P$ & 7.46$\pm$0.21 & Assumed 0 & 7.69$\pm$0.21 & Assumed 0 \\
$E_V$ & 2.37$\pm$0.19 &$\delta_{E_VT_V} =(-110 \pm 4)^\c$ & 1.11$\pm$0.22 &
 $\delta_{E_VT_V} =(-130 \pm 10)^\c$ \\
$C_V$ & 3.46$\pm$0.18 &$\delta_{C_VT_V} =(172 \pm 3)^\c$ & 4.05$\pm$0.17 &
 $\delta_{C_VT_V} =(162 \pm 4)^\c$ \\ \hline\hline
\end{tabular}
\end{center}
\end{table}

\section{$D \to P V$ amplitudes in the flavor-symmetric approach}

We list the $D^0 \to P V$ amplitudes appearing in Dalitz plots
of interest for the present discussion in
Tables \ref{tab:amps19} (for $\thet = 19.5^\circ$) and \ref{tab:amps11} (for
$\thet = 11.7^\circ$), including their representations and values in terms of
flavor SU(3) amplitudes.

\begin{table}
\caption{Amplitudes for $D^0 \to PV$ decays corresponding to Dalitz plots
of interest for the present discussion (in units of $10^{-6}$). Here we have
taken $\thet=19.5^\circ$.
\label{tab:amps19}}
\begin{center}
\begin{tabular}{c c c c c c c} \hline \hline
Dalitz & $D^0$ final & Amplitude & \multicolumn{4}{c}{Amplitude $A$} \\
 plot  & state  & representation & Re & Im & $|A|$ & Phase ($^\circ$) \\ \hline
&$\rho^+ \pi^-$ & --$\lambda(T_P+E_V)$ & --1.533 &  0.513 & 1.616 & 161.5 \\
$D^0 \to \pi^0 \pi^+ \pi^-$&$\rho^- \pi^+$ & --$\lambda(T_V+E_P)$ & --0.875 &  0.677 & 1.106 & 142.3 \\
&$\rho^0 \pi^0$ & $\frac{\lambda}{2}(E_P+E_V-C_P-C_V)$ &~0.819 & --0.477 & 0.947& -30.2 \\ \hline
&$K^{*+} K^-$&  $\lambda(T_P+E_V)$ &  1.533 & --0.513 & 1.616 & --18.5 \\
$D^0 \to \pi^0 K^+ K^-$&$K^{*-} K^+$   &  $\lambda(T_V+E_P)$ &  0.875 & --0.677 & 1.106 & --37.7 \\
&$\phi \pi^0$   &  $\lambda\,C_P/\s$ & --0.756 & --0.246 & 0.795 & --162 \\ \hline
& $\rho^+ K^-$ & $T_P+E_V$ & 6.649 & --2.227 & 7.012 & --18.5 \\
$D^0 \to K^- \pi^+ \pi^0$ &$K^{*-} \pi^+$& $T_V+E_P$ & 3.796 & --2.936 & 4.799 & --37.7 \\
&$\oks \pi^0$ & $(C_P-E_P)/\s$ & --3.173 & 1.010 & 3.33 & 162.3 \\
\hline \hline
\end{tabular}
\end{center}
\end{table}

\begin{table}
\caption{Same as Table \ref{tab:amps19} except with $\thet=11.7^\circ$.
\label{tab:amps11}}
\begin{center}
\begin{tabular}{c c c c c c c} \hline \hline
Dalitz &$D^0$ final & Amplitude & \multicolumn{4}{c}{Amplitude $A$} \\
 plot  & state  & representation & Re & Im & $|A|$ & Phase ($^\circ$) \\ \hline
&$\rho^+ \pi^-$ & --$\lambda(T_P+E_V)$ & --1.608 & 0.196 & 1.620 & 173.1 \\
$D^0 \to \pi^0 \pi^+ \pi^-$& $\rho^- \pi^+$ & --$\lambda(T_V+E_P)$ & --0.875 &  0.677 & 1.106 & 142.3 \\
&$\rho^0 \pi^0$ & $\frac{\lambda}{2}(E_P+E_V-C_P-C_V)$ & 0.879 & --0.407 & 0.968 & --24.8 \\ \hline
&$K^{*+} K^-$ & $\lambda(T_P+E_V)$ & 1.608 & --0.196 & 1.620 & --6.9 \\
$D^0 \to \pi^0 K^+ K^-$&$K^{*-} K^+$ & $\lambda(T_V+E_P)$ & 0.875 & --0.677 & 1.106 & --37.7 \\
&$\phi \pi^0$ & $\lambda\,C_P/\s$ & --0.756 & --0.246 & 0.795 & --162 \\ \hline
&$\rho^+ K^-$   &  $T_P+E_V$ & 6.977 & --0.850 & 7.028 & --6.9 \\
$D^0 \to K^- \pi^+ \pi^0$ & $K^{*-} \pi^+$ & $T_V+E_P$ & 3.796 & --2.936 & 4.799 & --37.7 \\
& $\oks \pi^0$ & $(C_P-E_P)/\s$ & --3.173 & 1.010 & 3.33 & 162.3 \\ \hline \hline
\end{tabular}
\end{center}
\end{table}

The amplitude representations in these tables imply the following interesting
relationships between the amplitudes:
\bea
\ca(D^0\to\rho^+\pi^-) =& - \ca(D^0\to K^{*+}K^-) &= \lambda
 \ca(D^0\to\rho^+K^-) \label{eqn:2}\\
\ca(D^0\to\rho^-\pi^+) =& - \ca(D^0\to K^{*-}K^+) &= \lambda
 \ca(D^0\to K^{*-}\pi^+) \label{eqn:3}
\eea
The above relationships are based only on SU(3) flavor symmetry and imply a
relative phase of $0^\circ$ or $180^\circ$ between amplitudes, independent of
the fitted parameter values. Taking ratios of these we obtain a ratio with
the same magnitude and phase independent of which Dalitz plot analysis one
chooses to look at:
\beq
R\,e^{{\it i}\Phi}=\frac{\ca(D^0\to\rho^+\pi^-)}{\ca(D^0\to\rho^-\pi^+)} =
\frac{\ca(D^0\to K^{*+}K^-)}{\ca(D^0\to K^{*-}K^+)} = \frac{\ca(D^0\to \rho^+K^-)}
{\ca(D^0\to K^{*-}\pi^+)}, \label{eqn:4}
\eeq
where $R$ and $\Phi$, respectively, denote the magnitude and phase of the
ratio. In the following section we calculate this ratio using Dalitz plot fits
and compare it with the predictions from the SU(3)-flavor-symmetric approach.

\section{Comparison of Dalitz plot data from experiments with predictions}

The representations of $D^0\to P V$ amplitudes mentioned in the previous
section do not contain the information about the vector meson decay to a pair
of light pseudoscalars. The fit fractions for an intermediate process $D^0 \to
R C$ in a $D^0\to A B C$ Dalitz plot ($A$, $B$ and $C$ are light pseudoscalars;
$R$ is the intermediate light vector resonance $AB$), however, also include the
fraction of the process $R\to A B$. The fraction of the vector meson's decay to
a pair of pseudoscalars can be simply represented by the corresponding isospin
Clebsch-Gordan factor.

It is important to note that the spin part of the amplitude for the process
$D^0 \to R C \to A B C$ is given by $T = -2\vec p_A\cdot\vec p_C$ where
$\vec p_i$ is the 3-momentum of the particle $i$ in the resonance rest frame,
implying that the phase of the corresponding amplitude changes by $\pi$ if we
switch the order of the daughters in the vector decay. The conventions used in
the present analysis are the same as those previously used for $D^0 \to \pi^0
\pi^+\pi^-$ in Ref.\ \cite{Bhattacharya:2010id, RApc, Mishra}, for $D^0 \to
\pi^0 K^+ K^-$ in Ref.\ \cite{Bhattacharya:2010ji, RApc, Mishra}, and for
$D^0 \to K^- \pi^+ \pi^0$ in Eq.\ (12) of Ref.\ \cite{Kopp:2000gv}. The
relevant Clebsch-Gordan factors are noted alongside the respective index
conventions in Table \ref{tab:clebsch} where we use Ref.\ \cite{Nakamura:2010}
for appropriate sign conventions.

\begin{table}
\caption{Conventions for order of the two pseudoscalars in vector meson decay
\cite{Bhattacharya:2010id,Bhattacharya:2010ji, Kopp:2000gv}.
\label{tab:clebsch}}
\begin{center}
\begin{tabular}{c c c c c c} \hline \hline
Dalitz Plot & \multicolumn{2}{c}{Bachelor Particle} & \multicolumn{3}{c}{Vector Meson Decay} \\
       & Meson & Index & Process & Indices & Clebsch factor \\ \hline
       & $\pi^0$ & 1 & $\rho^0   \to \pi^+ \pi^-$     & 23 & $1$     \\
$D^0 \to \pi^0 \pi^+ \pi^-$  & $\pi^+$   & 2 & $\rho^- \to \pi^0 \pi^-$ & 13
 & $1$    \\
   & $\pi^-$   & 3 & $\rho^+ \to \pi^0 \pi^+$   & 12 & -- $1$  \\ \hline
   & $\pi^0$ & 1 & $\phi   \to K^+ K^-$     & 23 & $1/\s$     \\
$D^0 \to \pi^0 K^+ K^-$  & $K^+$   & 2 & $K^{*-} \to K^- \pi^0$ & 31
 & --$1/\st$    \\
   & $K^-$   & 3 & $K^{*+} \to \pi^0 K^+$   & 12 & $- 1/\st$  \\ \hline
           & $K^-$   & 1 & $\rho^+ \to \pi^+ \pi^0$ & 23 & $1$ \\
$D^0 \to K^- \pi^+ \pi^0$  & $\pi^+$ & 2 & $K^{*-} \to K^- \pi^0$ & 13
 & --$1/\st$ \\
 & $\pi^0$ & 3 & $\oks \to K^- \pi^+$ & 12 & --$\sqrt{2/3}$  \\ \hline \hline
\end{tabular}
\end{center}
\end{table}

\begin{table}
\caption{Amplitudes for $D^0 \to PV$ decays from Dalitz plots of interest for
the present discussion (in units of $10^{-6}$). Here we have taken $\thet=19.5
^\circ$, and $\lambda = 0.2305$ \cite{BM}. The experimental amplitudes have
arbitrary overall normalization.
\label{tab:ampphase}}
\begin{center}
\begin{tabular}{c c c c c c} \hline \hline
$D^0$ final & Amplitude & \multicolumn{2}{c}{Theory \cite{Bhattacharya:2008ke,
Bhattacharya:2010id, Bhattacharya:2010ji}}
 & \multicolumn{2}{c}{Experiment * Clebsch factor} \\
 state & representation & Amplitude & Phase ($^\circ$) & Amplitude
 & Rel. Phase ($^\circ$) \\ \hline
$A(\rho^+ \pi^-)$ & $-\lambda(T_P + E_V)$ & 1.616$\pm$0.060
 & 161.5$\pm$1.6 & 0.823$\pm$0.004 & 0(def.)  \cite{Gaspero:2010pz}    \\
$A(\rho^- \pi^+)$ & $-\lambda(T_V + E_P)$ & 1.106$\pm$0.033 & 142.3$\pm$1.5
 & 0.588$\pm$0.008 & -- 2$\pm$0.6$\pm$0.6  \\
$A(\rho^0 \pi^0)$ & $\frac{\lambda}{2}(E_P + E_V$ &0.947$\pm$0.036
&--30.2$\pm$2.1&0.512$\pm$0.012&--163.8$\pm$0.6$\pm$0.4  \\
                  & $- C_P - C_V)$ &&&&\\ \hline
$A(K^{*+} K^-)$ & $\lambda(T_P+E_V)$ & 1.616$\pm$0.060 & --18.5$\pm$1.6
 & 1(def.) & 0(def.) \cite{Aubert:2007dc}   \\
$A(K^{*-} K^+)$ & $\lambda(T_V+E_P)$ & 1.106$\pm$0.033 & --37.7$\pm$1.5
 & 0.601$\pm$0.016 & --37.0$\pm$1.9$\pm$2.2 \\
$A(\phi \pi^0)$ & $\lambda\,C_P/\s$ &0.795$\pm$0.023&--162$\pm$1
&0.69$\pm$0.02&159.3$\pm$13.6$\pm$9.3 \\ \hline
$\lambda A(\rho^+ K^-)$ & $\lambda(T_P+E_V)$ & 1.616$\pm$0.060 &
--18.5$\pm$1.6 & 1(def.) & 0(def.) \cite{Kopp:2000gv} \\
$\lambda A(K^{*-} \pi^+)$ & $\lambda(T_V+E_P)$ & 1.106$\pm$0.033
 & --37.7$\pm$1.5 & 0.631$\pm$0.015$^a$ & --13.3$\pm$2.0$^a$ \\
 & & &            & 0.725$\pm$0.050$^b$ & --17.0$\pm$5.8$^b$ \\
$\lambda A(\oks \pi^0)$ &$\frac{\lambda}{\s}(C_P-E_P)/$ & 0.768$\pm$0.033
 &  162.3$\pm$1.7 & 0.457$\pm$0.036$^a$ &  172.8$\pm$2.2$^a$ \\
 & & &            & 0.494$\pm$0.011$^b$ &  179.8$\pm$8.0$^b$ \\ \hline \hline
\end{tabular}
\end{center}
\leftline{$^a$ Data from ``3-resonance fit'' of Ref.\ \cite{Kopp:2000gv}.}
\leftline{$^b$ Data from ``final fit'' of Ref.\ \cite{Kopp:2000gv}.}
\end{table}

Making use of the appropriate isospin Clebsch-Gordan factors listed in Table
\ref{tab:clebsch}, we may now calculate the magnitudes and phases of the
amplitudes listed in Eqs.\ (\ref{eqn:3}) and (\ref{eqn:4}). This is done in
Table \ref{tab:ampphase} for $\thet = 19.5^\circ$, where we also present the
corresponding results from SU(3) flavor-symmetry, side-by-side for easy
comparison.  In Table \ref{tab:ampphase} we also quote the magnitudes and
phases of other amplitudes from the relevant Dalitz plots for the sake of
completeness.  In order to obtain the experimental amplitudes to compare with
theory, we use the corresponding fit fractions from Refs.\
\cite{Gaspero:2010pz, Aubert:2007dc, Kopp:2000gv} and normalize them so as to
set the larger of the two amplitudes to equal $1$. In case the two processes
involved in this normalization have different values for the momentum $p^*$
then we also include a factor of the ratio $p^{*3}_A/p^{*3}_B$ where $A$ and
$B$ are the two processes involved in appropriate order. This takes account of
unequal phase space factors between the two processes.  For the $D^0 \to K^-
\pi^+ \pi^0$ Dalitz plot we use the data from both the ``3-resonance
fit'' and the ``final fit'' mentioned in Ref.\ \cite{Kopp:2000gv}.

In Table \ref{tab:ratio} we compare the magnitude $R$ and the phase $\Phi$ of
the ratio in Eq.\ (\ref{eqn:4}) obtained from the three different Dalitz plots
with the predictions from the SU(3) flavor-symmetric analysis. It is important
to notice that the agreement between theory and experiment on the value of $R$
is expected since the SU(3) flavor-symmetric approach makes use of some these
experiments. The fact that the corresponding relative phases agree fairly well
with each other, however, is working evidence for the flavor-symmetric approach.
A similar exercise when performed for $\thet = 11.7^\circ$ does not give any
significant changes and hence is omitted from this discussion.

\begin{table}
\caption{Comparison between predicted and measured ratios in Eq.\
(\ref{eqn:4}). Inputs were taken from Table \ref{tab:ampphase} above.
\label{tab:ratio}}
\begin{center}
\begin{tabular}{c c c c c} \hline \hline
Amplitude & \multicolumn{2}{c}{Predicted} & \multicolumn{2}{c}{Measured} \\
 ratio    & Magnitude ($R$) & Phase ($\Phi^\circ$) & Magnitude
 & Phase ($\Phi^\circ$) \\ \hline
$A(\rho^+\pi^-)/A(\rho^-\pi^+)$ & 1.461$\pm$0.070 & 19.2$\pm$2.2
 & 1.400$\pm$0.020 & 2.0$\pm$0.8 \\
$A(K^{*+}K^-)/A(K^{*-}K^+)$ & 1.461$\pm$0.070 & 19.2$\pm$2.2
 & 1.664$\pm$0.043 & 37.0$\pm$2.9 \\
$A(\rho^+K^-)/A(K^{*-}\pi^+)$ & 1.461$\pm$0.070 & 19.2$\pm$2.2
 & 1.585$\pm$0.037$^a$ & 13.3$\pm$2.0$^a$ \\
 & & & 1.379$\pm$0.100$^b$ & 17.0$\pm$5.8$^b$ \\ \hline \hline
\end{tabular}
\end{center}
\leftline{$^a$ Data from ``3-resonance fit'' of Ref.\ \cite{Kopp:2000gv}.}
\leftline{$^b$ Data from ``final fit'' of Ref.\ \cite{Kopp:2000gv}.}
\end{table}

\section{Comparison with previous analyses}

The results quoted in Tables \ref{tab:ampphase} and \ref{tab:ratio} are worth
comparing with our earlier results quoted in Refs. \cite{Bhattacharya:2010id,%
Bhattacharya:2010ji}.  The first three rows in Table \ref{tab:ampphase}
involve amplitudes and phases from the $D^0 \to \pi^0\pi^+\pi^-$ Dalitz plot.
These results agree with the findings of Ref.\ \cite{Bhattacharya:2010id}. The
relative amplitudes obtained using flavor SU(3) compare very well with those
found from experiment.  There is, however, some discrepancy between the
relative phases obtained experimentally and from theory. If we set the phase
for $A(\rho^+ \pi^-)$ to zero, then we obtain a relative phase discrepancy of
about $17^\circ$ for $A(\rho^-\pi^+)$ and $28^\circ$ for $A(\rho^0\pi^0)$.  The
S-wave interference contributions in the $D^0 \to \pi^0\pi^+\pi^-$ are
negligible. One possibility for the discrepancy is the existence of other
Dalitz plot solutions with relative phases closer to the flavor-SU(3)
predictions. In spite of this apparent discrepancy in relative phases, however,
the flavor SU(3) technique was able to reproduce branching fractions for the
isospin $(I = 0, 1, 2)$ amplitudes in $D^0 \to \pi^0\pi^+\pi^-$
\cite{Bhattacharya:2010id} similar to those seen in experiments by BaBar
\cite{Gaspero:2010pz}. This indicates that the flavor SU(3) technique is
successful in capturing some of the essential physics.

The amplitudes and phases from the Dalitz plot for $D^0 \to \pi^0 K^+ K^-$ are
quoted in rows four through six of Table \ref{tab:ampphase}. These results were
discussed in more detail in Ref.\ \cite{Bhattacharya:2010ji}.  Once again there
is good agreement relative amplitudes between experiment and flavor SU(3),
while relative phases don't agree so well. In this case we may set the phase
for $A(K^{*+} K^-)$ to zero.  This leads to a discrepancy in relative phases of
about $18^\circ$ for $A(K^{*-} K^+)$ and $57^\circ$ for $A(\phi \pi^0)$.
Note that the discrepancy is largest for $A(\phi \pi^0)$, for which we do not
have a cross-ratio relation.  In Ref.\ \cite{Bhattacharya:2010ji} we discussed
several sources for these discrepancies including inadequacy of the flavor-SU(3)
approach, the possibility of having other Dalitz plot fits with relative phases
closer to flavor-SU(3) predictions, and the need for proper parametrization of
the relevant S-wave $K\pi$ amplitudes.

In the remaining three rows of Table \ref{tab:ampphase} we calculate and
compare flavor SU(3) predictions of amplitudes and phases from $D^0 \to
K^-\pi^+\pi^0$ with experiment. As expected, the relative amplitudes from
theory agree fairly well with experiment. Let us set the phase for
$A(\rho^+ K^-)$ to zero to compare relative phases. We compare our results for 
relative phases with both the ``3-resonance fit'' as well as the ``final fit''
from Ref.\ \cite{Kopp:2000gv}. We find that the relative phases obtained from
flavor SU(3) for both $A(K^{*-}\pi^+)$ and $A(\oks \pi^0)$ agree with the
``final fit'' results and deviate from the ``3-resonance fit'' results by at
most $2.6 \sigma$. In this case the flavor-SU(3) approach seems to
produce results that are in acceptable agreement with experiment.

Finally in Table \ref{tab:ratio}, where we compute and compare ratios of the
amplitudes from Eq.\ (\ref{eqn:4}) that relate these three Dalitz plots, we
find that the amplitudes of these ratios agree very well with the results of
the corresponding experiments. There is, however, a residual discrepancy in
relative phases. Out of the three Dalitz plots considered for comparison this
phase discrepancy is almost the same for $D^0 \to \pi^0 K^+ K^-$ (about
$18^ \circ$) and $D^0 \to \pi^0 \pi^+ \pi^-$ (about $17^\circ$.)  In case of
the $D^0 \to K^- \pi^+ \pi^0$ Dalitz plot the relative phases agree with the
``final fit'' in Ref.\ \cite{Kopp:2000gv} while there is less than $2 \sigma$
deviation from the ``3-resonance fit'' result.  The results of Ref.\
\cite{Kopp:2000gv} are nearly ten years old.  An updated analysis of the $D^0
\to K^- \pi^+ \pi^0$ Dalitz plot could turn out to be useful in shedding more
light on the effectiveness of the flavor-SU(3) technique.

\section{Conclusion}

We have shown that several ratios of amplitudes and phases for $D^0 \to PV$
decays are predicted to have the same value.  The predictions of a
flavor-symmetric SU(3) analysis for this universal ratio agree fairly well with
results extracted from Dalitz plots for $D^0 \to \pi^+ \pi^- \pi^0$,
$D^0 \to \pi^0 K^+ K^-$, and $D^0 \to K^- \pi^+ \pi^0$.  Agreement is at
least as good as that for the process $D^0 \to K_S \pi^+ \pi^-$ compared
previously with $D^0 \to \pi^0 K^+ K^-$ \cite{Bhattacharya:2010ji}, with phase
discrepancies limited to $20^\circ$ or less.  (Agreement with magnitudes of
amplitudes is not surprising as the SU(3)-symmetric fits were based on observed
branching fractions.)
Cross ratios (independent of amplitude parametrizations) are shown to agree
with predictions better than parametrization-dependent predictions such as the
phase of the $D^0 \to \pi^0 \phi$ amplitude.  This could arise, for example, if
the cross ratios were less affected by flavor-SU(3) breaking than other
predictions of the flavor-SU(3) scheme.

The Dalitz plot analysis employed in our comparison for $D^0 \to K^- \pi^+
\pi^0$ is nearly ten years old \cite{Kopp:2000gv}.  While it yields reasonably
small errors in relative phases and magnitudes, considerably larger samples now
exist, thanks to BaBar, Belle, and CLEO.  It would be useful to update this
analysis in light of the new data.  Although S-wave $K \pi$ amplitudes play a
relatively small role in this process, it would be good to explore various
parametrizations of these amplitudes, as emphasized in
Ref.\ \cite{Bhattacharya:2010ji}, to be assured of the stability of the results.

\section*{Acknowledgements}

We thank C.-W. Chiang, M. Dubrovin, and B. Meadows for helpful discussions.
This work was supported in part by the United States Department of Energy through 
Grant No.\ DE FG02 90ER40560.

\end{document}